\documentclass{ws-ijmpe}

\def\mib#1{\mbox{\boldmath $#1$}}
\newcommand{\slp}{\raise.1ex\hbox{$/$}\kern-.63em\hbox{$p$}}
\newcommand{\slk}{\raise.15ex\hbox{$/$}\kern-.53em\hbox{$k$}}
\newcommand{\slpartial}{\raise.15ex\hbox{$/$}\kern-.53em\hbox{$\partial$}}

\begin{document}

\markboth{K. Watanabe and Y.Tsue}
{Amplification of Quantum Meson Modes in the Late Time of 
the Chiral Phase Transition}


\title{AMPLIFICATION OF QUANTUM MESON MODES\\
IN THE LATE TIME OF THE 
CHIRAL PHASE TRANSITION
}
\author{\footnotesize K.~WATANABE}

\address{Kochi Junior High School, Kochi Gakuen\\
Kochi 780-0956, Japan\\
Kazu.Watanabe@mb4.seikyou.ne.jp}

\author{\footnotesize Y.~TSUE
}

\address{Physics Division, Faculty of Science, Kochi University\\
Kochi 780-8520, Japan
\\
tsue@cc.kochi-u.ac.jp}

\maketitle


\begin{abstract}
We investigate the time evolution of the quantum meson modes 
in the late time of chiral phase transition.
In particular, it is shown that there exists a possible solution 
to the equation of motion for the quantum meson modes, 
which reveals a parametric resonance and/or resonance 
through forced oscillation induced by the small oscillation of 
the chiral condensate.
After that, we demonstrate 
the unstable regions for the quantum meson modes 
in both the cases of a uniform and spatially expanding system.
\end{abstract}

\section{Introduction}

In the experience of ultra-relativistic high energy nucleus-nucleus collision, 
it is concerned that chiral symmetry is restored.
However, 
when it is cooled down immediately, spontaneous symmetry breaking occurs again.
When chiral condensate returns to actual vacuum, 
the amplitude of quantum pion modes 
with low momenta even in the late time of chiral 
phase transition may grow. 
This phenomena may connect to the parametric resonance. 

In this paper, we are going to introduce the other possibility 
which may exist to this mechanism, 
especially we would like to focus on the possibility by forced oscillation.
To investigate the dynamical chiral phase transition, 
we use the variational approach to the O(4) linear sigma model 
in the Gaussian wave functional approximation.\cite{ref:VM,ref:TVM} 
It is shown that the amplification occurs in the mechanism of 
the resonance by forced oscillation as well as the parametric resonance 
induced by the small oscillation of the chiral condensate.\cite{ref:WT}

\section{TDVA with Gaussian wave functional}

In this section, we find the equations of motion for 
the chiral condensate and quantum fluctuations.
In order to treat the chiral condensate and the fluctuation modes around it 
self-consistently, 
we use 
the time-dependent variational approach (TDVA) with a Gaussian wave functional 
based on 
a functional Schrodinger picture.
We start with the following Hamiltonian, 
\begin{eqnarray}\label{2-1}
\hat{H} = \int d^3{\mib x} \left\{ \frac{1}{2} \pi ^2_a({\mib x}) 
+ \frac{1}{2} (\nabla \varphi_a({\mib x}))^2 + \frac{m^2}{2} \varphi _a({\mib x})^2 
+ \frac{\lambda}{24} (\varphi _a({\mib x})^2)^2 - h \varphi _0({\mib x}) 
\right\} 
, 
\end{eqnarray}
where $a$ runs from 0 to 3. The index 0 indicates the sigma field and 
1$-$3 indicate the pion fields.
The small $h$ means explicit chiral symmetry breaking term.
There are three model parameters used here, i.e., $m, \lambda$ and $h$. 
The model parameters $m^2$, $\lambda$ and $h$ are taken so as to 
reproduce the pion mass $M_\pi=138$ MeV, the sigma meson mass 
$M_0=500$ MeV and the pion decay constant $f_\pi(=\varphi_0)=93$ MeV 
in vacuum, 
explicitly, $m^2=-(519.70\ {\rm MeV})^2$, $\lambda=84.96$ and 
$h=(78.13\ {\rm MeV})^3$. 

We adopt the Gaussian wave functional as the trial wave functional 
in the framework of the functional Schr\"odinger picture.
The time-dependence of variational functions 
is governed by the time-dependent variational principle:
\begin{eqnarray}\label{2-5}
\delta \int dt 
\biggl\langle i \frac{\partial}{\partial t} - \hat{H} \biggl\rangle =0 \ . 
\end{eqnarray}
As the result, we obtain the following equations of motion:
\begin{eqnarray}
& &\left\{\partial_t^2 + M^2_0(t) 
- \frac{\lambda}{3} \bar{\varphi}_0(t)^2 \right\}
\bar{\varphi}_0(t) =h \ , 
\label{2-20}\\
& &\left\{ \partial_t^2 
+ {\mib k}^2 + M^2_{a}(t) \right\} 
u_a^{{\bf k}}(t) = 0 \ . 
\label{2-21}
\end{eqnarray}
These form a set of basic equations of motion for 
the chiral condensate and quantum meson fields.
The $\bar{\varphi}_0$ can be regarded as chiral condensate and 
the $u_a^{(n)}$ can be regarded as mode functions for quantum meson modes. 

We now demonstrate qualitatively 
the time evolution of the mean field $\bar{\varphi}_a$ 
and of the quantum meson mode functions $u_a^{(n)}$ in the case 
of a uniform system without spatially expansion.
We assume that $\bar{\varphi}_0 \neq 0$ and $\bar{\varphi}_i = 0$, with $i = 1\!~-\!~3$, 
that is, the chiral condensate points in the sigma direction.
In the numerical calculation, we adopt the box normalization with 
length $L$ in each direction. 
We then impose periodic boundary conditions for the fluctuation modes, 
namely, 
the allowed values of momenta are $k_x = (2 \pi/L)n_x$ and so on, 
where $n_x$ is an integer.
The fluctuation modes labeled by 
$(n_x,n_y,n_z)$ are included in each direction 
up to $n^2\equiv n_x^2+n_y^2+n_z^2 = 8^2$.
This corresponds to the three 
momentum cutoff $\Lambda \sim 1{\rm GeV}$~(more precisely, 990 MeV), 
since we have adopted 
the collisional region as $L^3 = (10\  {\rm fm})^3$.
Further, we assume that the fluctuation modes of the pion fields 
are all identical, and we denote them 
$u_1^{\bf k} = u_2^{\bf k} = u_3^{\bf k} \equiv u_{\pi}^{\bf k}$. 

\begin{figure}[t]
\centerline{\psfig{file=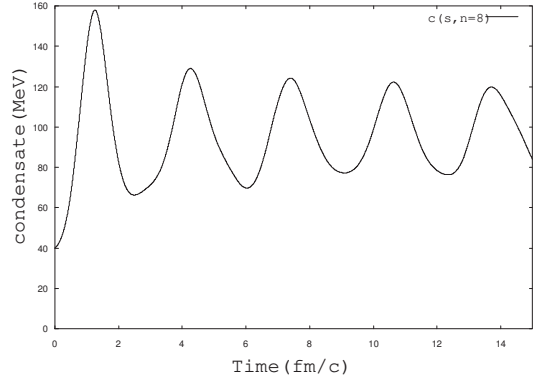,width=7cm}}
\vspace*{8pt}
\caption{The time evolution of the chiral condensate.
  		The horizontal and vertical axes represent time 
    		     and the value of the chiral condensation, respectively.}
\end{figure}
\begin{figure}[t]
\centerline{\psfig{file=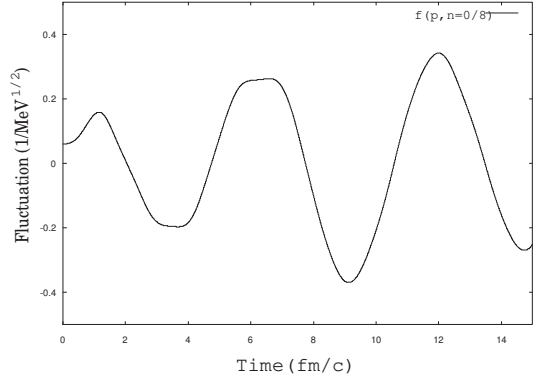,width=7cm}}
\vspace*{-28pt}
\caption{The time evolution of the fluctuation mode with $n=0$ 
                in the $\pi$-direction.}
\end{figure}
In Fig.~1, the time evolution of the chiral condensate is depicted. 
In order to avoid the complexity about problems of the initial 
conditions in relativistic heavy ion collisions, we only demonstrate 
the time evolution qualitatively with $\bar{\varphi}_0(t=0)=40$ MeV and 
$\dot{\bar{\varphi}}_0(t=0) = 0$. 
The reason we adopted the above initial conditions 
is that we are interested in the behavior of the time evolution of 
the chiral condensate and quantum meson modes in the late time of 
the dynamical chiral phase transition in this paper.
It is shown that the chiral condensate approaches the vacuum value with oscillation.

In Fig.~2, it is shown that 
the amplitude of the quantum pion mode with $n=0$ is amplified.
Similarly, the amplitude of the pion mode with $n=1$ is also amplified weakly. 
However, amplification is not realized for modes with $n \geq 2$. 
By contrast, the sigma meson modes with $n\ge 0$ are not amplified.

Thus, we conclude that, in this parameterization, the amplitudes of 
the lowest and first excited pion modes are amplified, but other 
quantum meson modes are not. 

The time evolution of the chiral condensate with the expanding 
geometry is also calculated.
It is seen that the damped oscillational behavior is realized more quickly 
than in the case without spatial expansion.

It should be noted here that the amplification occurs in the late time of 
the chiral phase transition in both the cases of no spatial 
expansion and one-dimensional spatial expansion. 
It is thus seen that, even when 
the chiral condensate oscillates around its vacuum value with small 
amplitude, there are amplified solutions of pion modes. 
The mechanism of the amplification is clarified in the next section.

\section{Late time of chiral phase transition}

In this section, 
we investigate the mechanism of the amplification for meson modes.
It is shown that one of possible mechanisms is a forced oscillation, 
as well as parametric resonance.

Let us investigate the time-dependent solutions around the static 
configurations.
The condensate and quantum meson modes can be expanded around the 
static solutions.
\begin{eqnarray}\label{3-2}
&&{\bar \varphi}_0(t) = {\varphi}_0 + \delta \varphi(t) \ , \nonumber\\
&&u_a^{\bf k}(t) = {u_a^{\bf k}}^s (t) + \delta u_a^{\bf k}(t) \ . 
\end{eqnarray}
Here, we consider the late time of the chiral phase transition. 
Then, $|\delta \varphi(t)|$ 
is small compared with the vacuum value, $\varphi_0$.
Further, we assume that $|\delta u_a^{\bf k}| \ll |{u_a^{\bf k}}^s|$. 
With the above approximation, we have following equation, 
\begin{eqnarray}\label{3-10}
\left[ \frac{d^2}{{dt'}^2} 
+ \omega_{\alpha}^2 [1-h_{\alpha}\cos(\gamma t') ] 
\right]\! 
{ \delta {\tilde u}_{\alpha}^{\bf k} }(t') 
= F_{\alpha}\! \cos(\gamma t') e^{-i \omega_{\alpha} t'} \ , 
\end{eqnarray}
where $\alpha=\sigma$ or $\pi$. 
Here, we introduce the dimensionless variables 
($\omega_{\alpha} , h_{\alpha} , F_{\alpha}$). 
If $F_{\sigma} \hspace{1mm}(F_{\pi})$ is negligible, then 
Eq.~(\ref{3-10}) is reduced to Mathieu's equation. 
In this case, the existence of the unstable solution for 
$\delta \tilde{u}_{a}^{\bf k}(t')$ may be expected. 
This phenomenon is well known as parametric resonance in classical mechanics.
On the other hand, 
forced oscillation may be realized through the effect of 
$F_{\sigma} \hspace{1mm}(F_{\pi})$,  even if the model parameters do not 
allow for unstable regions for the parametric resonance.
These phenomena are seen in the lowest and first excited pion modes, 
as demonstrated in the numerical calculations presented above. 

Let us search the the unstable regions for Eq.~(\ref{3-10}).
We perform the time integration over a certain interval for the magnitude 
of the quantum meson mode function, which means a kind of time average.
Then, we compare the time integral at a certain time step with 
a proceeding one.
If the value of this integral is larger than the last one for all time steps 
under consideration, then we decide that this mode is unstable.

Figure.3 shows the unstable region for quantum meson modes in 
the case of $F_\alpha = 3.0$.
The vertical axis represents $\omega_\alpha^2$ and the horizontal axis 
represents $\omega_\alpha^2 h_\alpha /2$.
The time integration is taken in 10 fm/$c$, in order to 
judge the stability of solution up to 100 fm/$c$.
It is shown that the unstable regions become wider due to 
the effect of the forced oscillation and/or the beat induced 
by the oscillation of the chiral condensate, as compared with the 
parametric resonance only. 

\begin{figure}[t]
\centerline{\psfig{file=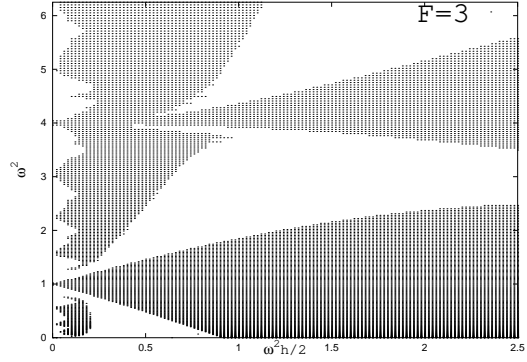,width=7cm}}
\vspace*{8pt}
\caption{The unstable regions for the solutions 
                in Eq.~(\ref{3-10}) with $F_{\alpha}=3.0$.}
\end{figure}
\begin{figure}[t]
\centerline{\psfig{file=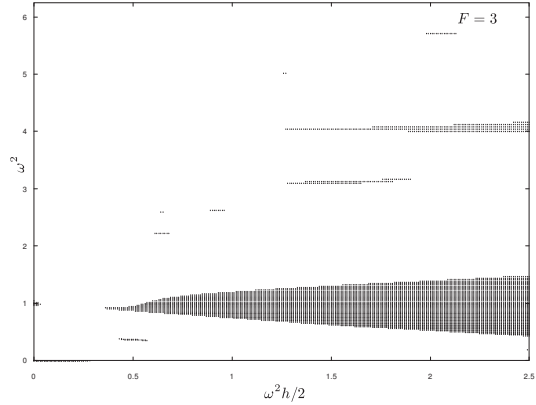,width=7cm}}
\vspace*{8pt}
\caption{The unstable regions for the solutions 
                in Eq.~(\ref{3-19}) with $F_{\alpha}=3.0$.}
\end{figure}

In the case with spatially expanding geometry, 
the time interval for the time integration to determine the unstable regions 
in our approach is taken as the rather short time 4 fm$/c$,  
because the unstable regions disappear 
when there is a strong damping effect for a long time. We calculate the time evolution 
up to 40 fm$/c$. 
The damping effect induced by the spatial expansion is introduced in the 
equations of motion for the quantum meson modes as follows: 
\begin{eqnarray}
\left\{ \frac{d^2}{{dt'}^2} + \omega_{\alpha}^2 [1-h_{\alpha} 
\cos(\gamma t') e^{- \mu t'} ] \right\} 
\delta {\tilde u}_{\alpha}^{[{\bf k}]}(t') 
= F_{\alpha} \cos(\gamma t') e^{- \mu t'} e^{-i \omega_{\alpha} t'} \ . \qquad
\label{3-19}
\end{eqnarray}
In the case of $F_\alpha=0$, the result for the unstable regions is 
well known, because this case corresponds to the case of the 
Mathieu equation with a friction term. 
Here, in Fig.4, we demonstrate the case of $F_\alpha = 3$ in Eq.~(7). 
In this case, the forced oscillation and beat may realize. 
However, these effects only give a small modification. 
The reason for this is that the friction has a strong effect in this model. 
As a result, the amplification of the quantum meson modes with the spatially 
expanding geometry is not strong, although the amplification appears. 
These results imply that the effect of the forced oscillation or the beat 
ceases to be effective for the spatially expanding geometry.

\section{Summary}

We have demonstrated that the amplitude of quantum pion modes are amplified 
even in the late time of the chiral phase transition 
through the mechanism of forced oscillation or the beat, 
as well as the parametric resonance, 
in the framework of the O(4) linear sigma model.
Also, we have determined the parameter regions 
in which the unstable solutions for the quantum meson mode functions exist.
Further, we have found that the effect of forced oscillation ceases 
due to the strong friction in the case with the spatially expanding geometry.

Of course, it is necessary to know the initial conditions in order to 
judge whether the amplification of the quantum meson mode occurs or not, 
that is, 
whether the parameters are in the unstable region or not, 
in realistic relativistic heavy ion collisions. 

\section*{Acknowledgements}

This work was partially
supported by Grants-in-Aid from the Japanese Ministry of Education, 
Culture, Sports, Science and Technology [Nos.15740156 and 18540278 (Y.T.)].



\begin{thebibliography}{0}



\bibitem{ref:VM} D. Vautherin and T. Matsui, {\it Phys. Rev.} 
{\bf D55} (1997) 4492.\\
D. Vautherin and T. Matsui, {\it Phys. Lett.} {\bf B437} (1998) 137.
\bibitem{ref:TVM} Y. Tsue, D. Vautherin and T. Matsui, 
{\it Prog. Theor. Phys.} {\bf 102} (1999) 313. \\
Y. Tsue, D. Vautherin and T. Matsui, {\it Phys. Rev.} {\bf D61} (2000) 076006.
\bibitem{ref:WT} K.~Watanabe, Y.~Tsue and S.~Nishiyama, 
{\it Prog. Theor. Phys.} {\bf 113} (2005) 369.

\end{thebibliography}
\end{document}